\documentclass[twocolumn,PRB,aps,psfig,preprintnumbers,superscriptaddress]{revtex4-2}
\usepackage{graphicx}
\usepackage{epstopdf}
\usepackage{float}
\usepackage{color}
\usepackage{amsmath}
\usepackage{ulem}

\begin{document}

\title{Enhancement of antiferromagnetic spin fluctuations in UTe$_2$ under pressure revealed by  $^{125}$Te NMR }

\author{Devi Vijayan Ambika}
\affiliation{Ames National Laboratory, U.S. DOE, Ames, Iowa 50011, USA}
\affiliation{Department of Physics and Astronomy, Iowa State University, Ames, Iowa 50011, USA}
\author{Qing-Ping Ding}
\affiliation{Ames National Laboratory, U.S. DOE, Ames, Iowa 50011, USA}
\author{Corey E. Frank}
\affiliation{Department of Physics, Maryland Quantum Materials Center, University of Maryland, College Park, Maryland 20742, USA}
\affiliation{NIST Center for Neutron Research, National Institute of Standards and Technology, Gaithersburg, Maryland 20899, USA}
\author{Sheng Ran}
\affiliation{ Department of Physics, Washington University, St. Louis, Missouri 63130, USA}
\author{Nicholas P. Butch}
\affiliation{Department of Physics, Maryland Quantum Materials Center, University of Maryland, College Park, Maryland 20742, USA}
\affiliation{NIST Center for Neutron Research, National Institute of Standards and Technology, Gaithersburg, Maryland 20899, USA}
\author{Yuji Furukawa}
\email{furukawa@ameslab.gov}
\affiliation{Ames National Laboratory, U.S. DOE, Ames, Iowa 50011, USA}
\affiliation{Department of Physics and Astronomy, Iowa State University, Ames, Iowa 50011, USA}

\date{\today}

\begin{abstract}
Characterizing magnetic fluctuations is one of the keys to understanding the origin of superconductivity in the spin-triplet superconductor UTe$_2$ which exhibits two superconducting (SC) phases (SC1 and SC2) under pressure: SC1 where a superconducting transition temperature of $T_{\rm c}$ decreases with pressure while $T_{\rm c}$ of SC2 rises with pressure. 
 Previously, D. Ambika {\it et al}. [Phys. Rev. B {\bf105}, L220403 (2022)] have reported the possible coexistence of ferromagnetic (FM) and antiferromagnetic (AFM) spin fluctuations in UTe$_2$ under pressure from their nuclear magnetic resonance (NMR) measurements.
To delve the relationship between the magnetic fluctuations and the two SC phases,  we have carried out detailed $^{125}$Te NMR measurements on a single crystal of UTe$_2$ with $T_{\rm c}$ = 1.6 K  at various pressures ranging from 0 to 2.05 GPa.
 By comparing the temperature $T$ dependence of nuclear spin-lattice relaxation rates divided by temperature 1/$T_1T$  with that of the Knight shift $K$ for magnetic fields along the $a$, $b$, and $c$ directions, we evidence the enhancement of AFM spin fluctuations with increasing pressure.
  Based on the results, we suggest that FM spin fluctuations are more favorable for SC1 and AFM spin fluctuations are crucial for SC2.
  Our findings will inspire further study on this material to understand the peculiar SC phases in detail.

\end{abstract}

\maketitle

\section{Introduction}

\begin{figure}[h!tb]
\includegraphics[width=1.0\columnwidth]{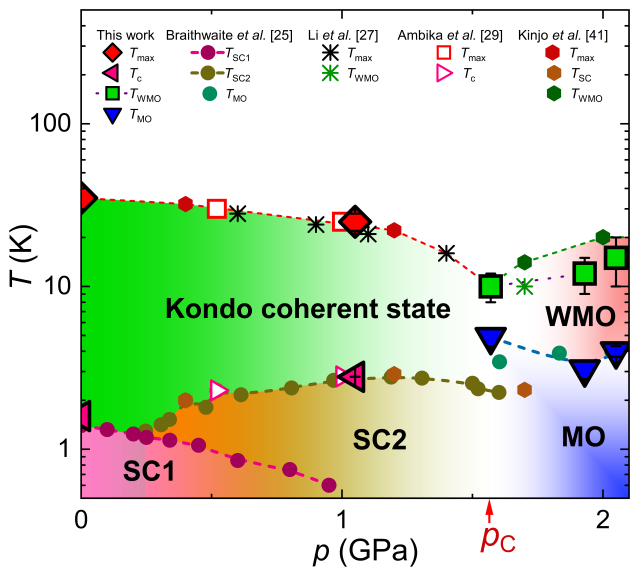}
\caption{Pressure - temperature ($p$-$T$) phase diagram of UTe$_2$.  $T_{\text{max}}$ is the temperature at which a broad maximum is observed in the temperature dependence of the magnetic susceptibility or Knight shift $K$ under $H \parallel b$ reported from Refs. \cite{Li2021,Kinjo2022,Ambika2022}, $T_{\text{c}}$ is the superconducting critical temperature from Refs. \cite{Braithwaite2019,Kinjo2022,Ambika2022}, $T_{\text{WMO}}$ is the weakly magnetic ordered temperature from Refs. \cite{Kinjo2022,Li2021}, $T_{\text{MO}}$ is the magnetic ordering temperature from Refs. \cite{Li2021,Kinjo2022}.  The slightly larger filled symbols are the corresponding temperatures from the present NMR measurements (see text). It is noted that, although the SC1 phase inside the SC2 phase, shown in the figure, has been reported to be different from SC1 from the NMR measurements (named as SC3) \cite{Kinjo20232},  we write it with one SC phase (SC1) for simplicity.}
\label{fig:Fig1}
\end{figure}

    The discovery of superconductivity in UTe$_2$  with a superconducting transition temperature $T_{\rm c}$ of $\sim$ 1.6 K  \cite{Ran2019,Aoki2019} and now raised up to $\sim$ 2.1 K \cite{Sakai2022,Rosa2022}  has triggered intense research activity to characterize the physical properties of the compound, as it has been suggested to be a possible candidate of spin-triplet superconductors \cite{Ran2019, Aoki2019, Aoki2022, Knebel2019}. 
      The spin-triplet superconducting (SC) state is suggested from a highly anisotropic upper critical field $H_{c2}$ that exceeds the Pauli limit \cite{Ran2019, Aoki2019, Aoki2022, Knebel2019}, and also from a tiny reduction of the Knight shift ($K$) below $T_{\rm c}$ in all three crystallographic directions  \cite{Nakamine2019, Nakamine2021, Nakamine20212, Fujibayashi2022, Matsumura2023, Kinjo2023, Kinjo20232}. 
    In addition, recent phase-sensitive measurements utilizing the Josephson effect have also supported the spin-triplet superconducting state \cite{Zixuan2025}. 
    
     One of the important key parameters to characterize materials, especially for the mechanism of superconductivity,  is magnetic fluctuation.     
      Ferromagnetic (FM) fluctuations have been initially considered to play an important role in the triplet pairing,  as suggested by  nuclear magnetic resonance (NMR) measurements \cite{Tokunaga2019, Tokunaga2023} where Ising-like FM fluctuations along the $a$ axis were reported. 
      At the same time, antiferromagnetic (AFM) spin fluctuations with the incommensurate wave-vector of $q$ = (0, 0.57, 0) was also revealed by neutron scattering (NS) measurements \cite{Knafo2021, Duan2020, Duan2021, Raymond2021}, suggesting the importance of the AFM spin fluctuations as well as the FM fluctuations, although  more recent NS measurements suggest interband spin excitons arising from $f$-electron hybridization as a possible origin of the magnetic excitations \cite{Butch2022,Halloran2025}.
      Recent NMR measurements also pointed out the existence of both FM and AFM fluctuations at ambient pressure \cite{Fujibayashi2023,Matsumura2025}.

     UTe$_2$ also exhibits many intriguing properties under pressure. 
      With the application of pressure,  two SC phases appear around 0.25 GPa \cite{Braithwaite2019, Thomas2020}. 
      As depicted in Fig. \ref{fig:Fig1},  the SC phase with lower $T_{\rm c}$ (SC1) is suppressed continuously with increasing pressure.
 In contrast,  the $T_{\rm c}$  of the SC2 phase is enhanced,  takes a maximum $T_{\rm c}$ $\sim$ 3 K at around 1.2 GPa, and then the SC2 phase is suppressed rapidly at higher pressures around $p_{\rm c}$~$\sim$~1.5~GPa. 
     Above the critical pressure $p_{\rm c}$, an AFM magnetic ordered phase appears \cite{Thomas2020,Li2021, Knafo2025} where two phases were suggested: weakly magnetic order (WMO) state below  $T_{\rm WMO}$ and long-range magnetic order state below $T_{\rm MO}$ as illustrated in Fig. \ref{fig:Fig1}.

     Previously, we reported, from NMR measurements on a single crystal of UTe$_2$ at pressures from 0 to 1.57 GPa under magnetic field $H$ parallel to the $b$ axis, a possible coexistence of FM and AFM spin fluctuations under pressure, with the AFM spin fluctuations becoming enhanced with increasing pressure \cite{Ambika2022}.  Recent theoretical studies also point out that the interplay between these spin fluctuations is pivotal in understanding the emergence of multiple SC phases \cite{Tei2024,Hakuno2024}, therefore, it is crucial to investigate the evolution of spin fluctuations in detail under varying pressure conditions.

     In this paper, we carried out  $^{125}$Te NMR measurements on a $^{125}$Te-enriched single crystal of UTe$_2$ under magnetic fields $H$ parallel to the three crystalline axes $a$, $b$, and $c$ at various pressures ranging from 0 to 2.05 GPa to investigate the details of how the magnetic fluctuations change with pressure. 
    From the results of nuclear spin-lattice relaxation rate divided by temperature 1/$T_1T$ measurements under the $H$ $\parallel$ $a$, $b$, and $c$ directions in comparison with the Knight shift $K$ results, we found the enhancement of AFM spin fluctuations under pressure where the hyperfine field fluctuations associated with the AFM fluctuations are found to mainly enhance along the $a$ and $c$ axes at the Te sites.
  Based on the analysis of the data, we suggest that FM spin fluctuations are more favorable for SC1 and AFM spin fluctuations are crucial for SC2. 

  \begin{figure}[h!tb]
\centering
\includegraphics[width=0.6\columnwidth]{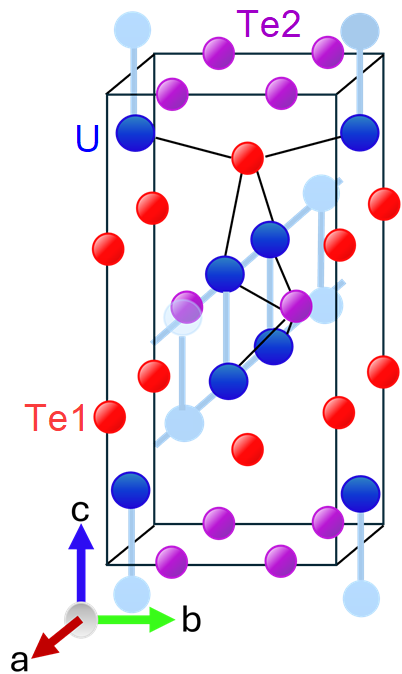}
\caption{ Crystal structure of UTe$_2$. The U atoms (blue circles) form a two-leg ladder structure  along the $a$ axis. 
    Two inequivalent tellurium (Te) sites are indicated as Te1 (red circles) and Te2 (violet circles). The light blue circles representing the U atoms are added to show the two-leg ladder structure.}
\label{fig:Fig2}
\end{figure}

\section{Experimental Details}

\begin{figure*}[h!tb]
\centering
\includegraphics[width=2\columnwidth]{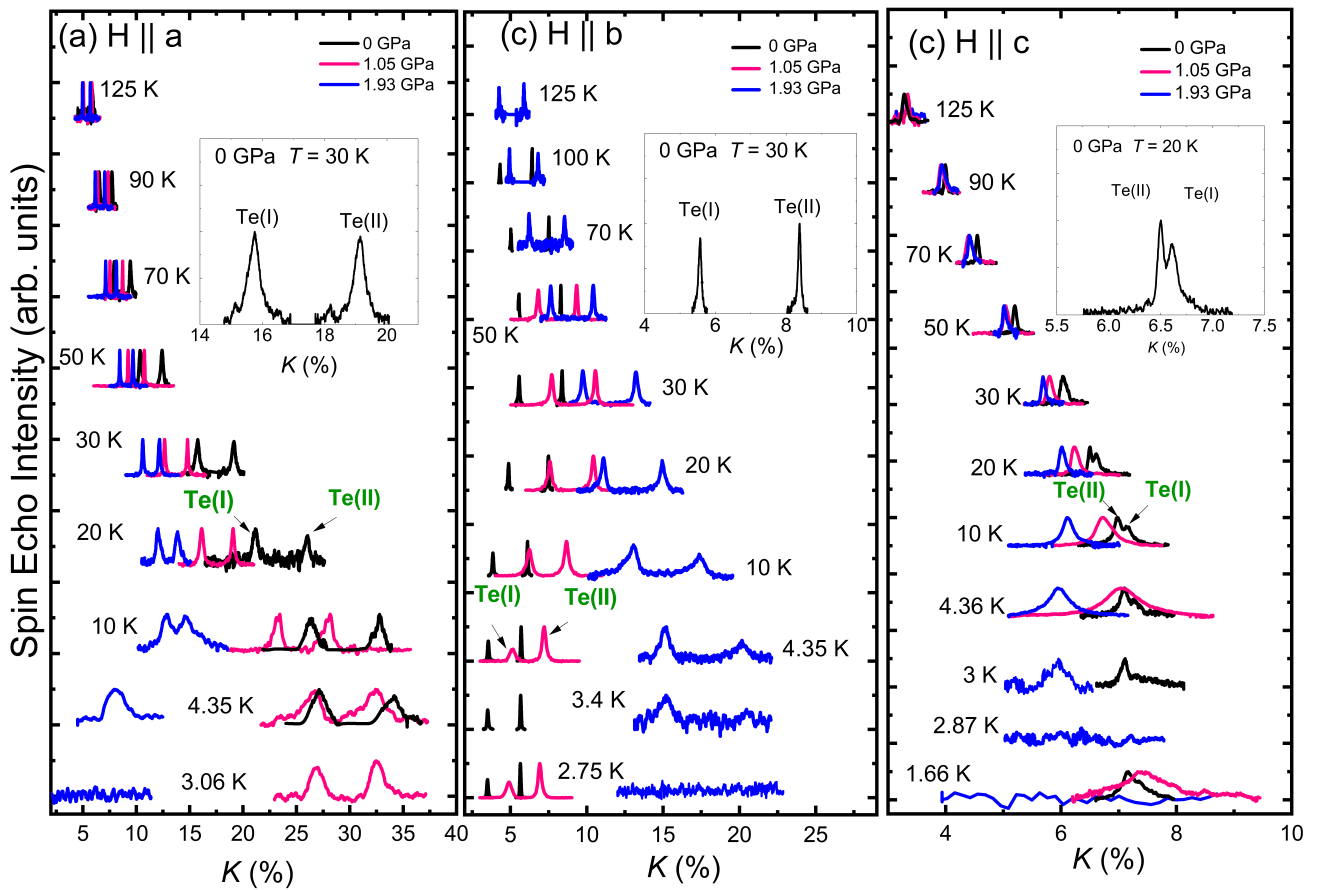}
\caption{ (a)-(c)  Temperature dependence of the field-swept $^{125}$Te NMR spectra measured on the UTe$_2$ single crystal for (a) $H$ $\parallel$ $a$  under pressure (0 GPa at $f$ = 50.35 MHz, 1.05 GPa at $f$ =12.2 MHz, and 1.93 GPa at $f$ =12.2 MHz), (b) $H$ $\parallel$ b (0 GPa at $f$ = 41 MHz, 1.05 GPa at $f$ =  12.2 MHz, and 1.93 GPa at $f$ = 55.6 MHz, and (c) $H$ $\parallel$ $c$ (0 GPa , 1.05 GPa, and 1.93 GPa at $f$ = 50.35 MHz, respectively). 
The horizontal axis is Knight shift $K$ defined by $K$ = ($H_0$ -- $H$)/$H$ where $H_0$ = 2$\pi$$f$/$\gamma_{\rm N}$,  $f$ is NMR resonance frequency, and $H$ is the external magnetic field.}
\label{fig:Fig3}
\end{figure*}

    The $^{125}$Te-enriched single crystal of  UTe$_2$ with $T_{\rm c}$ = 1.6 K (at $H$ = 0 T) at ambient pressure was synthesized by the chemical vapor transport method using iodine as the transport agent \cite{Ran2019}. 
   NMR measurements of $^{125}$Te (nuclear spin $I$ = $\frac{1}{2}$, gyromagnetic ratio $\frac{\gamma_{\rm N}}{2\pi}$ = 13.454 MHz/T) nuclei were conducted using a laboratory-built phase-coherent spin-echo pulse spectrometer for the pressures 0, 1.05, 1.93, and 2.05 GPa with a NiCrAl/CuBe piston-cylinder cell using  Daphne 7373 as the pressure-transmitting medium which guarantees high hydrostaticity up to 2.2 GPa at room temperature \cite{Yokogawa2007}.
     The calibration of pressure was accomplished by $^{63}$Cu nuclear quadrupole resonance in Cu$_2$O \cite{Fukazawa2007, Reyes1992} at 77 K.  
     The superconducting transition temperature $T_{\rm c}$ = 2.8 K for 1.05 GPa at zero magnetic field ($H$ = 0) is determined in the AC susceptibility measurements using an NMR tank circuit. 
     The $^{125}$Te-NMR spectra were obtained by sweeping $H$ at fixed NMR frequencies ($f$) under $H$ parallel to the $a$, $b$, and $c$ axes where the misalignment of the crystal is expected to be less than a few degrees. The 1/$T_{\rm 1}$ was measured with a saturation recovery method.  
      $1/T_1$ at each $T$ was determined by fitting the nuclear magnetization $M$ versus time $t$  using the exponential function $1-M(t)/M(\infty) = e^{-(t/T_{1})}$ for $^{125}$Te NMR,  where $M(t)$ and $M(\infty)$ are the nuclear magnetization at $t$ after the saturation and the equilibrium nuclear magnetization at $t$ $\rightarrow$ $\infty$, respectively. 
 The NMR data for $H \parallel b$ at $p$ = 0 and 1.57 GPa are taken from those reported in our previous paper \cite{Ambika2022}. The uncertainty in the NMR shift and $T_{\rm 1}$ was estimated by analyzing the maximum and minimum bounds of scattering of the data for each measurement. Furthermore, uncertainties for derived quantities were determined using error propagation methods.
 
   UTe$_2$ crystallizes in a body-centered orthorhombic structure with the $Immm$ space group \cite{Hutanu2020}, where the U atoms form a two-leg ladder structure with legs along the $a$ axis and rung along the $c$ axis as shown in Fig.~\ref{fig:Fig2} \cite{VESTA}. 
    There are two crystallographic inequivalent Te sites occupying $4j$ and $4h$ sites with local symmetries $mm2$ and $m2m$, respectively. 
     Following the previous paper \cite{Tokunaga2019}, these sites are denoted by Te1 and Te2. 
The Te1 site is located within the distorted tetrahedron formed by two first and two second nearest-neighboring U atoms which belong to three different ladders.
On the other hand, Te2 is surrounded by the four nearest neighbor U atoms forming a square-like structure within a ladder.    
     The NMR spectra exhibit two distinct NMR lines, expected from the two Te sites, Te1 and Te2 \cite{Tokunaga2019}. 
      The specific assignment of these lines to each Te site has remained to be determined \cite{Fujibayashi2023}. 
      Thus we designate the peaks as Te(I) and Te(II), following the nomenclature used in the previously published paper \cite{Tokunaga2019,Fujibayashi2023}. 
However, in this paper we argue that  Te(I) and Te(II) are most likely assigned to the Te1 and Te2, respectively, in the crystal structure as reported in Supplemental Material (SM) \cite{SM}.   
This is the same with the initial site assignment of Te NMR lines in Ref. \cite{Tokunaga2019}, and Fujibayashi \textit{et al}. also pointed out such a possibility in Ref. \cite{Fujibayashi2023}.

\section{Results and Discussion}

\begin{figure*}[h!tb]
\includegraphics[width=2.1\columnwidth]{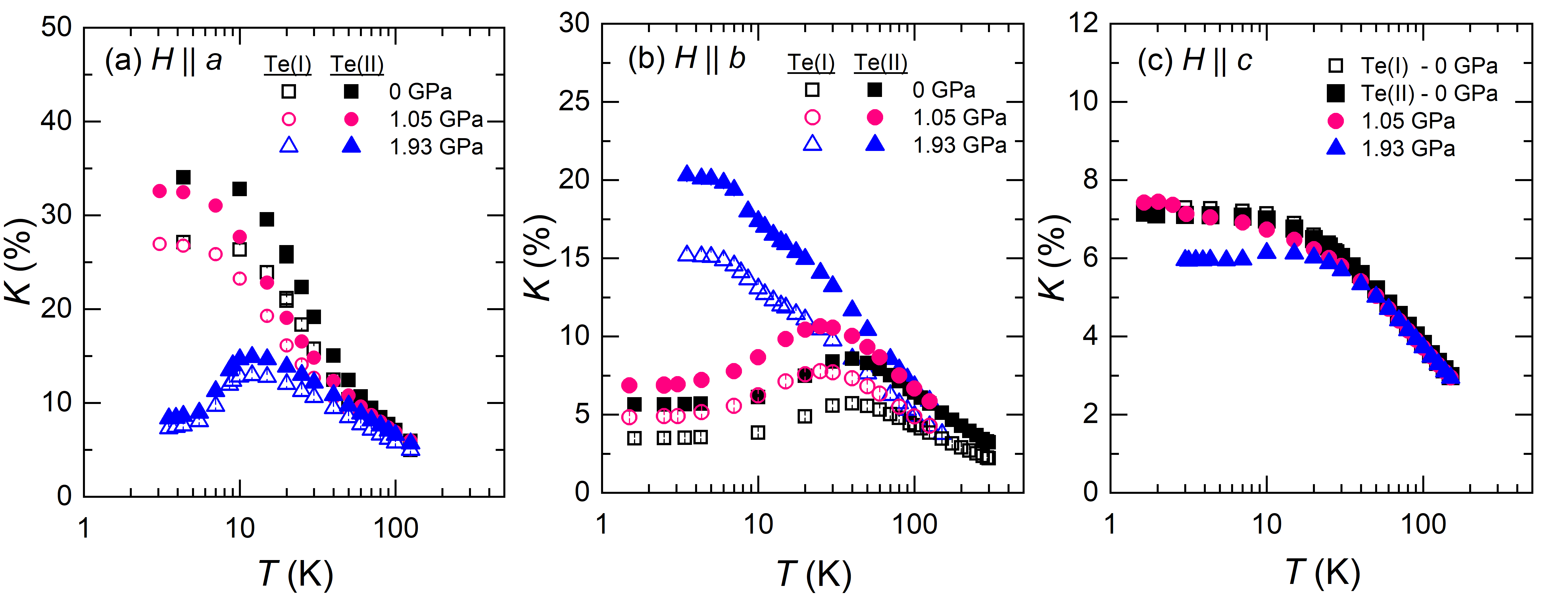}
\caption{$T$ dependences of $K$s for both Te(I) and Te(II) under different pressures  (ambient pressure in black, 1.05 GPa in pink, and 1.93 GPa in blue) under  $H$ parallel to all three crystalline axes: (a) $H$ $\parallel$ $a$, (b) $H$ $\parallel$ $b$, (c) $H$ $\parallel$ $c$. }
\label{fig:Fig4}
\end{figure*}

\subsection{$^{125}$Te NMR spectrum}

 Figures~\ref{fig:Fig3}(a), \ref{fig:Fig3} (b), and \ref{fig:Fig3}(c)  show the temperature dependence of the $H$-swept $^{125}$Te-NMR spectra of a single crystal  UTe$_2$ for three different magnetic field directions parallel to the $a$ ($H$ $\parallel$ $a$), the $b$ ($H$ $\parallel$ $b$) and the $c$ ($H$ $\parallel$ $c$) axes, respectively,  under different pressures of ambient, 1.05 GPa, and 1.93 GPa.
  The horizontal axes are Knight shift $K$ defined by $K$ = ($H_0$ -- $H$)/$H$ where $H_0$ is Larmor field given by 2$\pi$$f$/$\gamma_{\rm N}$ with $f$ being the NMR resonance frequency. 

     At ambient pressure,   two distinct $^{125}$Te-NMR lines for  $H$ $\parallel$ $a$ and $H$ $\parallel$ $b$  and two partially overlapped peaks for $H$ $\parallel$ $c$ are observed, consistent with the previous reports  \cite{Tokunaga2019, Fujibayashi2023}. 
      Following the previous NMR papers  \cite{Tokunaga2019, Fujibayashi2023}, the two lines with smaller  and greater $K$ for $H$ $\parallel$ $a$ and $H$ $\parallel$ $b$  are denoted to Te(I) and Te(II), respectively, and the line with the larger $K$ value under  $H$ $\parallel$ $c$ is assigned to Te(I),  as shown in Figs.~\ref{fig:Fig3}(a), \ref{fig:Fig3}(b) and \ref{fig:Fig3}(c).
    At 1.05 GPa below the critical pressure $p_{\rm c}$ $\sim$ 1.5 GPa, similar spectra, but slightly different $K$ values have been observed.

   In contrast, at $p$ = 1.93 GPa above $p_{\rm c}$, although similar spectra with slightly different $K$s were observed at higher temperatures above $\sim$40 K, quite different temperature dependencies of NMR spectra and $K$ were observed below that temperature. 
   In particular, no signals were observed below $\sim$ 3 K for all $H$ directions. 
  This is due to the pressure-induced long-range magnetic ordering above $p_{\rm c}$ \cite{Li2021,Kinjo2022, Knafo2025}. 
   Furthermore, we also notice that,  for $H$ $\parallel$ $b$, the signal intensity of Te(II) starts decreasing below $\sim$ 10 K, and only the signal of Te(I) can be well observed at 3.4 K as shown in Fig.~\ref{fig:Fig3}(b). 
   This could be due to a short-range (or weakly) magnetic ordered state. The intensity difference between the two Te sites we observe for 0 GPa and 1.05 GPa for $H$ $\parallel$ $b$ is confirmed to be due to the difference in the nuclear spin-spin relaxation rate.
   We will discuss the temperature dependence of the NMR spectra for $H$ $\parallel$ $b$ above $p_{\rm c}$ in more detail below.

\begin{figure*}[h!tb]
\centering
\includegraphics[width=2.05\columnwidth]{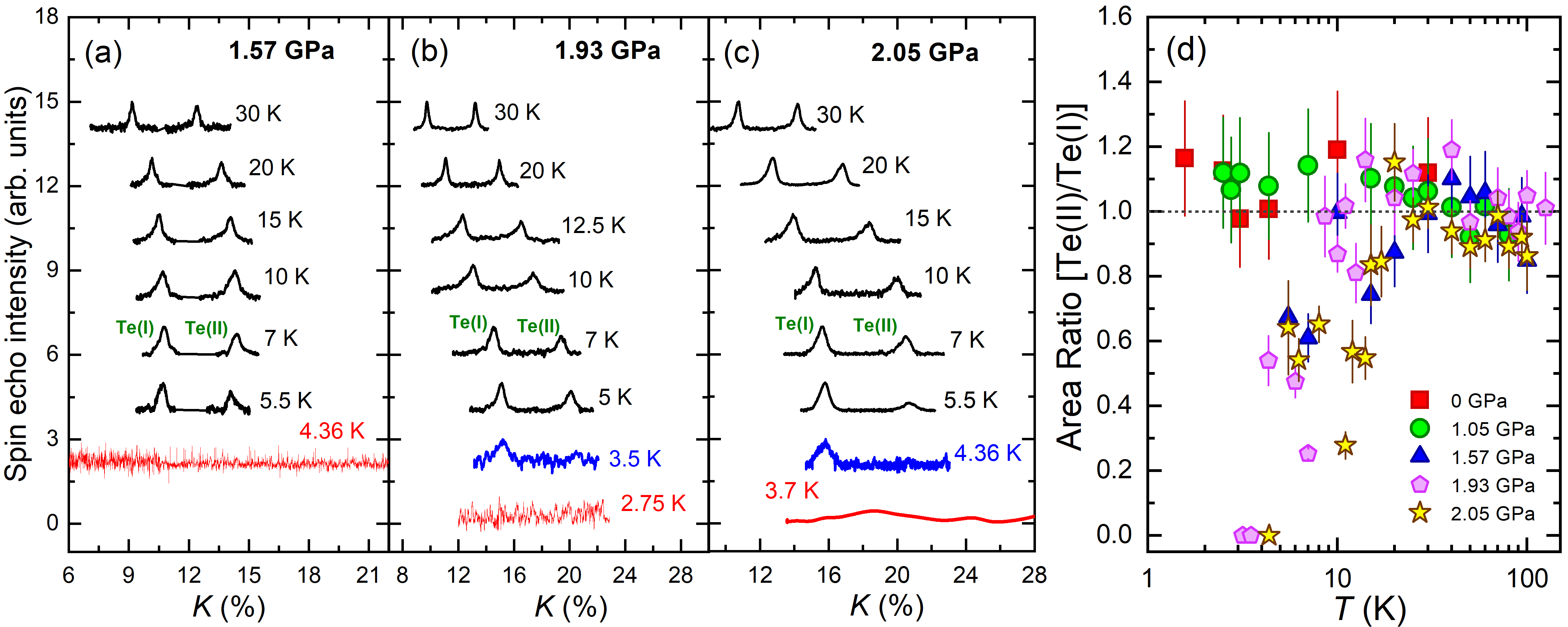}
\caption{Temperature dependence of the $H$-swept $^{125}$Te-NMR spectra of a single crystal  UTe$_2$ for (a) $p$ = 1.57 GPa, (b) 1.93 GPa, and (c) 2.05 GPa for $H$ $\parallel$ $b$ where the horizontal axis is Knight shift $K$.  For $p$ = 1.57 and 2.05 GPa, the spectra were measured at $f$~=~50.35~MHz ($H_0$ =  3.74~T).  $f$~=~55.60~MHz ($H_0$~=~4.132~T) was used at $p$ = 1.93 GPa. (d) Temperature dependence of the signal intensity ratio of Te(II) to Te(I)  for $p$ = 0, 1.05, 1.57, 1.93 and 2.05 GPa. Here the data for 1.57 GPa are from \cite{Ambika2022}.}
\label{fig:Fig5}
\end{figure*}

  Figure~\ref{fig:Fig4} depicts the temperature dependence of $^{125}$Te Knight shift $K$ for both Te sites for all the measured pressures and $H$ directions.
   The $K$ values of Te(II) are greater than those of Te(I) for the $a$ and $b$ axis directions. 
   This is due to the difference in hyperfine coupling constants, which are estimated to be $A_{aa}$(Te(I)) = 3.8 T/$\mu_{\rm B}$ and  $A_{aa}$(Te(II)) = 4.7 T/$\mu_{\rm B}$ for the $a$ axis, and $A_{bb}$(Te(I)) = 3.5 T/$\mu_{\rm B}$ and $A_{bb}$(Te(II)) = 5.2 T/$\mu_{\rm B}$ for the $b$ axis, respectively, at ambient pressure, from the so-called $K$-$\chi$ plot analysis (see SM  \cite{SM}). 
   In contrast, along the $c$ axis, the $K$ values of Te(I) and Te(II) are nearly the same, and $A_{cc}$ for Te(I) and Te(II) is estimated to be 4.1 T/$\mu_{\rm B}$  at the ambient pressure \cite{SM}. These coupling constants were also found to be nearly independent of $p$ \cite{Kinjo2022}.      
   
   The temperature dependence of $K$ for $H$ $\parallel$ $a$  at ambient pressure is similar to the results reported previously \cite{Tokunaga2022, Fujibayashi2023} where, with decreasing $T$, the  $K$ values increase and then level off at low temperatures below 10 K. 
    It was reported that the different $T$ dependence between $K$ and $\chi_a$ at low temperatures below $\sim$ 10 K is due to the U atom defects in low-$T_{\rm c}$ crystals \cite{Tokunaga2022} where $\chi$ increases with decreasing temperature due to the defect effects while $K$ reflecting the intrinsic behavior of $\chi$ shows a nearly constant behavior at low temperatures. With increasing $p$,  the values of $K$ are slightly suppressed at 1.05 GPa. 
   However, at 1.93 GPa above $p_{\rm c}$, a markedly different temperature dependence of $K$ was observed, exhibiting a maximum around 10 K, as shown in Fig.~\ref{fig:Fig4}(a). 
    This significant change in the $T$ dependence of $K$ under high $p$ is consistent with the magnetic susceptibility measurements \cite{Li2021}. 
   It is noted that the temperature of the broad maximum is close to $T_{\rm WMO} \approx 12$ K,  the onset of a weakly magnetic ordered (WMO) state reported in earlier studies \cite{Li2021,Kinjo2022}.
   
Similarly,  a drastic change in the $T$ dependence of $K$ for $H \parallel b$ under pressures can be seen in Fig.~\ref{fig:Fig4}(b).
While $K$ under $H$ $\parallel$ $b$ show broad maxima at $T_{\rm max}$ $\sim$ 35--40 K at ambient pressure and $T_{\rm max}$ $\sim$ 25 K at 1.05 GPa, consistent with the magnetic susceptibility data \cite{Ran2019, Aoki2019, Li2021},  the $K$ keeps increasing and saturates below $\sim$8 K at 1.93 GPa. 
These results suggest that the magnetic easy axis changes from the $a$ axis to the $b$ axis upon crossing $p_{\rm c}$, and is in good agreement with the $\chi$ data under pressure \cite{Li2021}.
 
    In the case of $H$ $\parallel$ $c$,  the application of pressure leads to only a moderate change in the temperature dependence of $K$, where a slight suppression of $K$ at 1.93 GPa is observed below $\sim$ 30 K showing a weak but visible broad maximum around 10-12 K which is close to $T_{\rm WMO} \approx 12$ K.  The above discussed values of $K$ for $H$ $\parallel$ $b$ and $H$ $\parallel$ $c$ under pressure are consistent with previously reported results \cite{Kinjo2022}.

\begin{figure*}[h!tb]
\centering
\includegraphics[width=2\columnwidth]{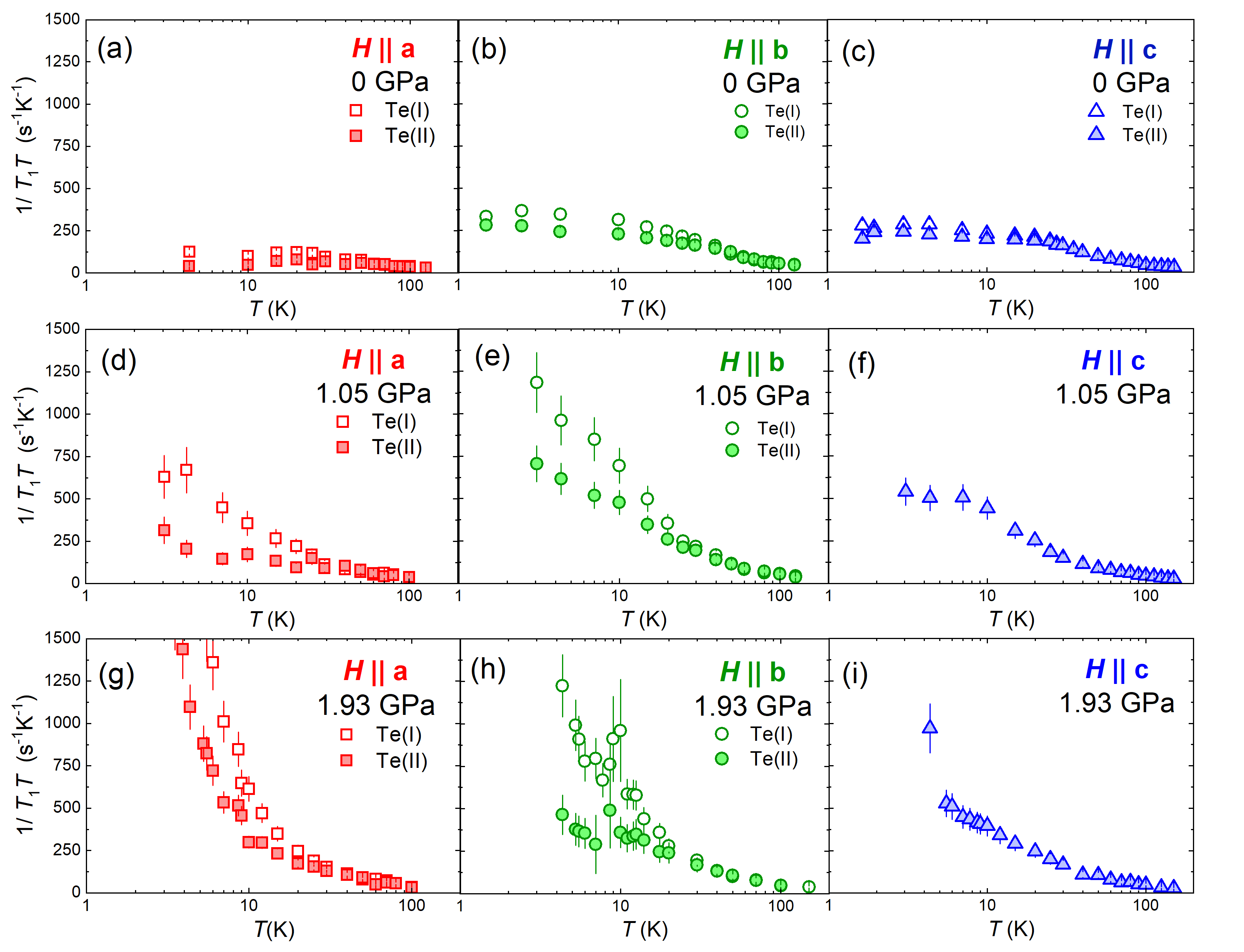}
\caption{Temperature dependence of 1/$T_1T$ for Te(I) (open symbols) and Te(II) (solid symbols) at $p$ = 0 GPa [(a), (b), (c)], 1.05 GPa [(d), (e), (f)], and  1.93 GPa [(g), (h), (i)]  under  $H$ $\parallel$ $a$  (in red),  $H$ $\parallel$ $b$ (in green),  and $H$ $\parallel$ $c$ (in blue). The data at ambient pressure under  $H$ $\parallel$ $b$ are from \cite{Ambika2022}. 
}
\label{fig:Fig6}
\end{figure*}

\subsection{ $^{125}$Te NMR spectra for $H$ $\parallel$ $b$ above $p_{\rm c}$}

    As described above, the NMR signal intensities of Te(I) and Te(II) exhibit different temperature dependencies under a pressure of 1.93 GPa at low temperatures, but this is not observed below $p_{\rm c}$.   
To further examine this signal-intensity change, we measured the NMR spectra under $H$ $\parallel$ $b$ at a different pressure of 2.05 GPa above $p_{\rm c}$. 
  The spectra are presented in Fig.~\ref{fig:Fig5}(c), together with the previous data at 1.57 GPa \cite{Ambika2022} and the spectra at 1.93 GPa shown in Figs.~\ref{fig:Fig5}(a) and \ref{fig:Fig5}(b), respectively.
  As described,  in the case of 1.93 GPa,  the NMR line of Te(II) begins to diminish below $\sim$10 K, and only the Te(I) line can be well observed at 3.5 K, and then both signals disappear between 2.75 and 3.5 K, as shown in  Fig.~\ref{fig:Fig5}(b) [also Fig.~\ref{fig:Fig2}(b)].
  A similar behavior is observed at 2.05 GPa where the Te(II) signal intensity starts decreasing below $\sim$10 K and both signals disappear between 3.7 and 4.36 K. 
   In contrast,  at 1.57 GPa, although the signal intensities of Te(II) slightly decrease with respect to that of Te(I) below 10 K, both signals remain present at 5.5 K, and both signals disappear at nearly the same temperature of 4.36 K.   
   As mentioned above, the disappearance of both signals is due to the pressure-induced long-range magnetic order above $p_{\rm c}$. 
   Based on these results, we estimate the long-range magnetic ordering temperature $T_{\rm MO}$ to be $\sim$ 4.9(6) K at 1.57 GPa, $\sim$ 3.1(4) K at 1.93 GPa, and  $\sim$4.0(3) K at 2.05 GPa.

  \begin{figure*}[h!tb]
\centering
\includegraphics[width=2\columnwidth]{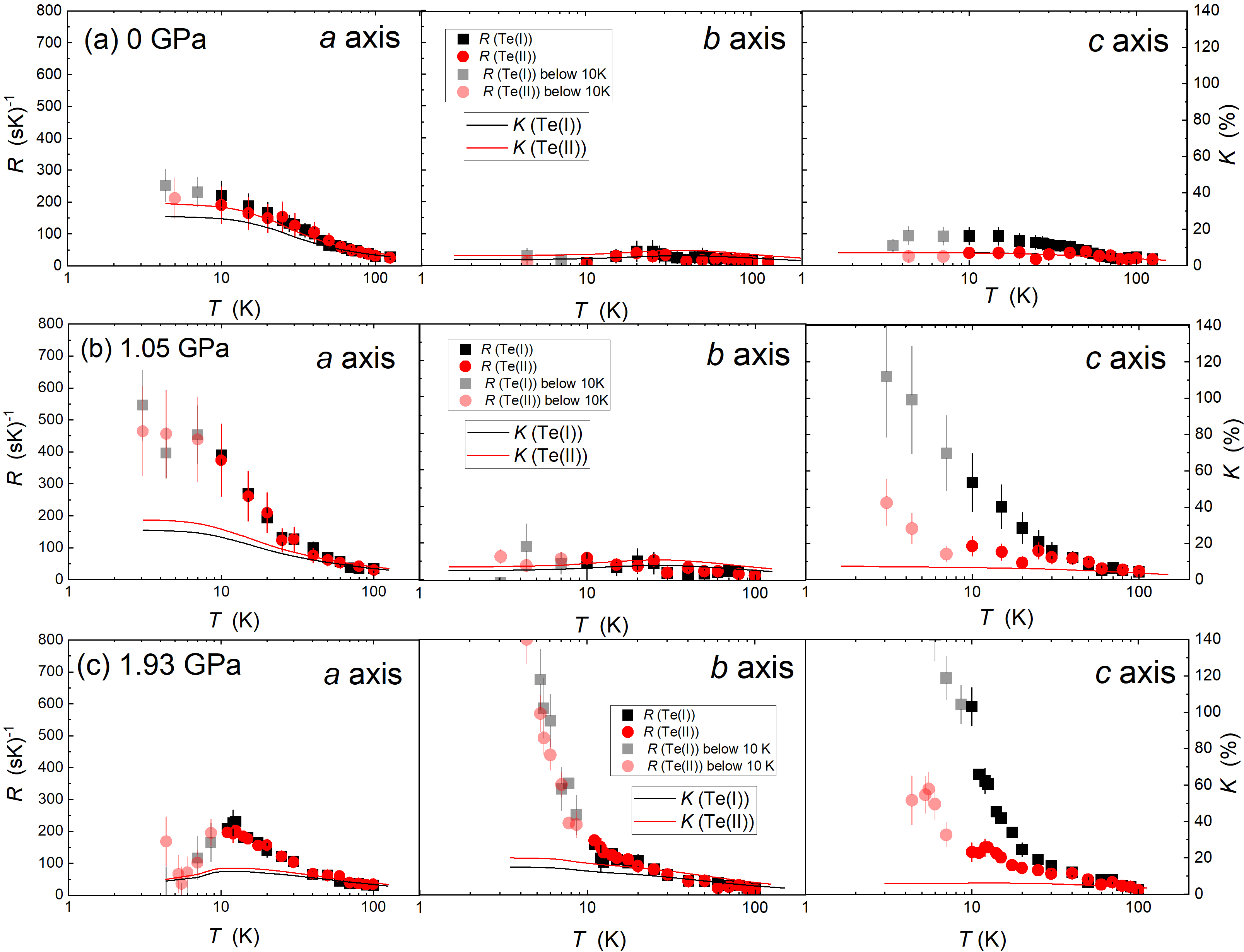}
\caption{ Temperature dependence of $R_{a}$ (left), $R_{b}$ (middle), and $R_{c}$ (right) for the $p$ = 0 GPa (a), 1.05 GPa (b), and 1.93 GPa (c) for Te(I) (black squares) and Te(II) (red circles), together with the temperature dependence of the corresponding $K$ of Te(I) (black curves) and Te(II) (red curves). The $R_{i}$ data below 10 K were also shown with translucent symbols for reference.}
\label{fig:Fig7}
\end{figure*}

   The disappearance of the Te(II) line above $T_{\rm MO}$ could be due to the short-range magnetic ordered state. To better understand the difference in the change of the signal intensities of the two signals, we plotted the ratio of the signal intensity of Te(II) to that of Te(I) as a function of temperature, where the signal intensities were corrected by the nuclear spin-spin relaxation time $T_2$. 
   As shown in Fig.~\ref{fig:Fig5}(d), the ratio clearly begins to decrease below $\sim$10(3) K for 1.57 GPa and 1.93 GPa and  $\sim$15(5) K for 2.05 GPa, unlike 0 GPa and 1.05 GPa where the ratio stays close to one. 
   Those temperatures are close to the value of $T_{\rm WMO}$ $\sim$ 9.5 K at 1.7 GPa reported from the magnetic susceptibility measurements \cite{Li2021}.

\subsection{$^{125}$Te spin-lattice relaxation rate 1/$T_1$}

   To shed the light on the evolution of the magnetic fluctuations with pressure, we measured the $^{125}$Te spin-lattice relaxation rate (1/$T_1$)  at the peak positions of the Te(I) and Te(II) lines at $p$ = 0, 1.05, and 1.93 GPa under three different magnetic field directions ($H$ $\parallel$ $a$, $H$ $\parallel$ $b$, and $H$ $\parallel$ $c$), whose results are summarized in Figs.~\ref{fig:Fig6}(a)-\ref{fig:Fig6}(i).  A general trend is that  1/$T_1T$ for both Te sites under all three magnetic field directions appears to increase with increasing pressure.  
  In particular, as reported before \cite{Ambika2022},  1/$T_1T$ for Te(I) enhances more than that of Te(II) under $H$ $\parallel$ $b$ at $p$ = 1.05 GPa, which was attributed to the enhancement of AFM spin fluctuations under pressure.   
   With the present new 1/$T_1T$ data, although we were not able to resolve the $T_1$ values for the two Te lines along the $c$ axis because of the overlapping of the two signals,  it is clearly evident that, for not only $H$$\parallel$$b$ but also for $H$$\parallel$$a$, 1/$T_1T$ for Te(I) enhances more than that of Te(II) under pressures of 1.05 and 1.93 GPa.  The values of the 1/$T_1T$ are consistent with the previously reported values at ambient pressure for both Te sites for all three magnetic field directions \cite{Fujibayashi2023} and under pressure for the Te(I) site with the magnetic field parallel to the $b$ and $c$ axes \cite{Kinjo2022}.

 We analyze the current $T_1$ data following the previous procedure \cite{Tokunaga2019}.
  By writing the directional magnetic fluctuation components, $R_{i,\alpha}$ where $i =$ I, II and $\alpha, \beta, \gamma $ = $\{a, b, c\}$ for each Te($i$) as
\begin{equation}
R_{i,\alpha} = \frac{\gamma_{\rm N}^2k_{\rm B}}{2}\sum_{q, \xi}|A_{{\rm {hf}}, i}^{\alpha\xi}(q)|^{2} \frac{\chi_\xi^{\prime \prime}\left(q, \omega_{\rm N}\right)}{\omega_{\rm N}},
\label{Eq1}
\end{equation}
$(1/T_1T)_{i,\alpha}$ can be described by  $(1/T_1T)_{i,\alpha}$ = $R_{i,\beta}+R_{i,\gamma}$ where $(\alpha,\beta,\gamma)$ are mutually orthogonal directions. 
Here $\chi^{\prime \prime}\left(q, \omega_{\rm n}\right)$ is the imaginary part of dynamic magnetic susceptibility at the NMR frequency ($\omega_{\rm N}$) along the $\xi$ (= $\{a, b, c\}$) directions and $A_{{\rm {hf}}, i}^{\alpha\xi}(q)$ is the $q$-dependent hyperfine coupling tensor for $^{125}$Te nuclei.
  Utilizing the 1/$T_1$ data measured under the three different $H$ directions, we derived the temperature dependence of $R_{i,\alpha}$ for each direction at three different pressures as shown in Figs.~\ref{fig:Fig7}(a)-\ref{fig:Fig7}(c).
   Here it is noted that a recent NMR study comparing  a $``$early stage$"$ $T_ {\rm c}$ =1.6 K crystal and $``$ultra-clean$"$  $T_ {\rm c}$ = 2.1 K crystal showed a difference in $T$ dependence of 1/$T_1T$ at low temperatures (mainly below $\sim$10 K) due to the U deficiency in the low $T_{\rm c}$ crystal \cite{Matsumura2025}.  
  Therefore, since the high-temperature (above $\sim$10 K) behavior seems to be consistent across the samples \cite{Matsumura2025}, we discuss the magnetic fluctuations based on our data only above 10 K  which are shown with the solid filled symbols in  Figs.~\ref{fig:Fig7}(a)-\ref{fig:Fig7}(c), together with the data below 10 K shown with the translucent symbols for reference. 
    In  the figures, we also plotted the temperature dependence of the corresponding $K$ of both the Te sites in each panel (the right axes of the figures).
   Since $K$ reflects the temperature dependence of the real component  $\chi^{\prime}(\vec{q}, \omega_0)$ with $q$ = 0 and $\omega_0$ = 0, the comparison between $R_{i,\alpha}$ and $K$ allows us to examine the $T$ dependence of $\sum_{q, \xi}|A_{{\rm {hf}}, i}^{\alpha\xi}(q)|^{2} \frac{\chi_\xi^{\prime \prime}\left(q, \omega_{\rm N}\right)}{\omega_{\rm N}}$ with respect to the uniform static susceptibility $\chi^{\prime}$(0, 0).

At ambient pressure ($p$ = 0 GPa),  the temperature dependence of $R_{i,\alpha}$ roughly scales with the corresponding $K$, suggesting that the $q$ = 0 ferromagnetic fluctuations are dominant.
It is noted that $R_{i,{\rm c}}$ for Te(I) is greater than those for Te(II) along the $c$ axis at low temperatures, as has been pointed out in Ref. \cite{Fujibayashi2023}. Furthermore, the temperature dependence of $R_{i,{\rm c}}$ for Te(I) does not scale with that of $K$, whereas for Te(II), both $R_{i,{\rm c}}$ and $K$ are scaled  (see SM \cite{SM}). These results indicate that Te(I) picks up the additional magnetic fluctuations other than $q$ = 0, providing evidence for the hyperfine field fluctuations with AFM nature along the $c$ axis at Te(I). It is also noted that, even though $K$ for Te(I) is smaller than that of Te(II) at low temperatures, $R_{i,{\rm a}}$ of Te(I) is slightly greater (or almost similar) than that of Te(II). 
This indicates a site dependence in magnetic fluctuations, consistent with the previous NMR report \cite{Matsumura2025}, suggesting that Te(I) detects the hyperfine field fluctuations associated with the AFM nature along the $a$ axis as well, even though the ferromagnetic fluctuations are dominant. 
The existence of the AFM fluctuations along the $a$ axis could be more clearly seen in the temperature dependence of  $R_{i,{\rm a}}$ divided by the corresponding $K$, $R_{i,{\rm a}}/K$, as shown in SM \cite{SM}.
It is noted that the magnetic fluctuations picked up at the Te sites originate from U spin fluctuations via hyperfine interactions. Since the Te1 and Te2 sites are located at different sites with distinct local environments, the differences in magnetic fluctuations detected at the Te(I) and Te(II) sites can be attributed to the distinct hyperfine coupling tensors (see SM \cite{SM}) where Te(I) and Te(II) are most likely assigned to the crystallographic Te1 and Te2 sites, respectively.

   At $p$ = 1.05 GPa, the significant enhancement of $R_{i,{\rm a}}$ and  $R_{i,{\rm c}}$ at 1.05 GPa can be seen in Fig.~\ref{fig:Fig7}(b), which can be attributed to an additional contributions of the AFM spin fluctuations under pressure. 
    Especially,  a clear difference in $R_{i,{\rm c}}$ between Te(I) and Te(II) is observed, evidencing that again Te(I) detects the AFM spin fluctuations more than Te(II).
  This is consistent with the previous NMR measurements under pressures up to 1.57 GPa \cite{Ambika2022}.
  It turns out that the different temperature dependence of 1/$T_1T$ of Te(I) and Te(II) for $H \parallel b$ under pressure reported before in Ref. \cite{Ambika2022} was due to the different temperature dependence of $R_{i,{\rm c}}$ for Te(I) and Te(II) because no difference in  $R_{i,{\rm a}}$ in the two Te sites was observed. 
   
\begin{figure*}[h!tb]
\includegraphics[width=1.25\columnwidth]{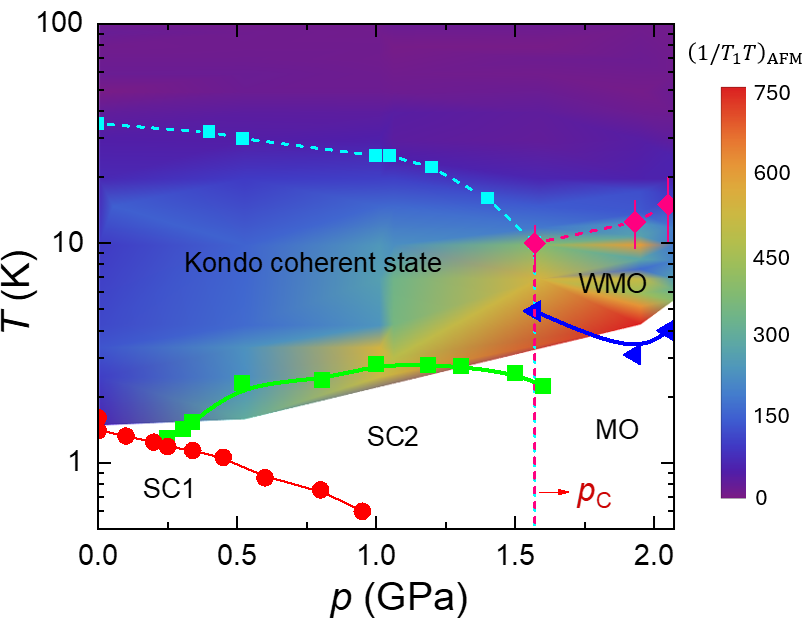}
\caption{ $p$-$T$ contour plot showing the magnitude of AFM spin fluctuations tentatively represented by $(1/{T_{1}T})_{\rm{AFM}}$  = $1/T_{1}T$(Te(I)) - $1/T_{1}T$(Te(II)) as defined in the text. The $1/T_{1}T$ data with $H \parallel b$ for 0 - 2.05 GPa are used to estimate $(1/{T_{1}T})_{\rm{AFM}}$. The $T_1$ data for $p$ = 0 GPa, 0.52 GPa, 1 GPa, and 1.57 GPa are from Ref. \cite{Ambika2022} and the $T_1$ data for $p$ = 2.05 GPa is given in SM \cite{SM}.  The various transition temperatures, including $T_{\rm{c}}$, $T_{\rm{max}}$, $T_{\rm{WMO}}$, and $T_{\rm{MO}}$ from Fig. 1 are also plotted. }
\label{fig:Fig8}
\end{figure*}  

        When $p$ is greater than the critical pressure $p_{\rm c}$,  although $R_{i,{\rm c}}$  for Te(I) and Te(II) is similar to that at 1.05 GPa showing the AFM spin fluctuations picked at the Te(I) site, it is clear that $R_{i,{\rm b}}$ for both Te sites is enhanced compared to the lower-pressure case below $p_{\rm c}$. 
       It is noted that, although the $K$ for the $b$-axis direction increases at this $p$, the enhancement of  $R_{i,{\rm b}}$ for both Te sites is greater than the increase in $K$. Therefore, AFM spin fluctuations are actually enhanced along the $b$ axis. 
       This is quite different from the lower-pressure case where the weak $b$-axis magnetic fluctuations were observed.  
       Although we do not have a clear idea to explain the large enhancement of AFM spin fluctuations along the $b$ axis at present, it could be related to the change in the magnetic easy axis from the $a$ to $b$ axes above $p_{\rm c}$ \cite{Li2021} or could be due to the different wave vector in the antiferromagnetically ordered state \cite{Knafo2025}. Further studies are needed to clarify the magnetic fluctuations above $p_{\rm c}$.   
      It is also noted that the $R_{i,{\rm a}}$ for both Te sites are slightly suppressed. 
      This suppression cannot be explained by the large reductions in $K$, again evidencing the hyperfine field fluctuations associated with AFM nature along the $a$ axis.

      Finally, it is important to mention that, as pressure increases, the $T_{\rm c}$ of SC2 increases while the $T_{\rm c}$ of SC1 decreases. 
     Given the observed enhancement of AFM spin fluctuations with pressure, the AFM spin fluctuations would play an important role in the appearance of SC2. 
      In this context, it is interesting to point out that recent theories proposed a possibility of the spin-triplet state induced by AFM spin fluctuations \cite{Kreise2022,Tei2024,Hakuno2024}. 
      Conversely, the AFM fluctuation would likely be unfavorable for SC1 state as the $T_{\rm c}$ of SC1 decreases upon pressure, which would suggest the importance of FM fluctuations for SC1.

   To qualitatively illustrate the correlation between AFM spin fluctuations and superconductivity in UTe$_2$, we present the evolution of AFM spin fluctuations under pressure in a contour plot (Fig.~\ref{fig:Fig8}).
   Here the magnitude of AFM spin fluctuations is tentatively represented by  $(1/{T_{1}T})_{\rm{AFM}}$  = $1/T_{1}T$(Te(I)) $-$ $1/T_{1}T$(Te(II)) using the 1/$T_1$ data measured under $H \parallel b$, which mainly corresponds to the difference in  $R_{\rm c}$ for the two Te sites. 
   Although it is better to use the quantities estimated from the comparison between the $R_{\alpha}$ and  $K_{\alpha}$ data for representing the AFM spin fluctuations, we here used only the 1/$T_1T$ data under  $H \parallel b$ to make the contour plot due to the limited number of data sets for $R_{\alpha}$ under pressure. Thus, the plot reflects the partial contributions of the AFM fluctuations. Nevertheless,  the contour plot captures the main feature of the enhancement of AFM fluctuations with increasing pressure, concomitant with the rise of $T_{\rm{c}}$ for SC2. It is noted that the small reduction in $T_{\rm{c}}$ close to $p_{\rm c}$ could be due to the long-range magnetic ordering appearing above $p_{\rm c}$. A similar suppression of $T_{\rm{c}}$ near the AFM phase has been observed in iron-based superconductors, such as FeSe$_{1-x}$S$_x$ under pressure \cite{Matsuura2017}. 
   As mentioned above,  $T_{\rm{c}}$ for SC1 is suppressed as pressure is augmented.  Given the FM spin fluctuations dominate at ambient pressure, these results would suggest that  FM spin fluctuations are more favorable for SC1 and AFM spin fluctuations are crucial for SC2, as pointed out above. A recent NMR study by Kinjo {\it et al}. reported  that  the \textit{\textbf{d}} vector has no $b$ component in SC2 while it exhibits a finite component along the $b$ axis in SC1 (and also SC3 defined in Ref \cite{Kinjo20232}).  
     It is interesting if these differences are related to the distinct magnetic fluctuations suggested for the different SC phases.  
Further studies are therefore highly requested to elucidate the origin of superconductivity as well as the SC properties in the two SC1 and SC2 phases.

\section{Summary}

We performed  $^{125}$Te NMR measurements on the $^{125}$Te-enriched single crystal of UTe$_2$ ($T_{\rm c}$ = 1.6 K) to investigate the evolution of magnetic fluctuations in UTe$_2$ under pressure. 
Following the previous NMR studies under pressures which suggested that the AFM spin fluctuations are enhanced with the application of pressure \cite{Ambika2022},  we measured 1/$T_1T$ under the magnetic field parallel to the three crystalline axes at various pressures ranging from 0 to 2.05 GPa.  
 As a result,  we evidenced the enhancement of AFM spin fluctuations under pressure. 
 In particular, from the directional resolved dynamical spin susceptibilities derived from the $1/T_1T$ data of the three magnetic field directions,  we found the hyperfine fluctuations at the Te sites associated the AFM spin fluctuations are mainly enhanced along the $a$ and $c$ axes, which is the origin of the enhancement of 1$/T_1T$ for $H$ $\parallel$ $b$ under pressure.
   It is also shown that the Te(I) site picks up the enhancements of hyperfine field fluctuations along the $c$ axis compared to the Te(II) site.
   From this difference, combined with the calculations of the hyperfine tensors for the two Te sites in the crystal (see SM \cite{SM}), we suggest that Te(I) and Te(II) denoted in the NMR lines are assigned to Te1 and Te2 denoted in the crystal, respectively.
 
    We also revealed the magnetically ordered state above the critical pressure $p_{\rm c}$ from the NMR measurements. 
   From the disappearance of the Te NMR lines, the long-range magnetic order temperatures were estimated to be $\sim$4.9(6) K at 1.57 GPa, $\sim$3.1(4) K at 1.93 GPa, and  $\sim$4.0(3) K at 2.05 GPa. 
   The short-range (or weakly) magnetic order was also detected by the different temperature dependence of two Te NMR lines below 10-15 K, which we attributed to the development of ferromagnetic correlations within the U ladders.     
 
Our results suggest that  AFM spin fluctuations  play an important role in the appearance in SC2, as $T_{\rm c}$ of SC2 increases with pressure.
On the other hand, as the SC1 phase is suppressed with pressure, the AFM spin fluctuations are unfavorable for the phase.          
    Therefore, we suggest that FM spin fluctuations are more favorable for SC1 and AFM spin fluctuations are crucial for SC2.
 Further studies are strongly requested to elucidate the origin of superconductivity of the two SC1 and SC2 phases, as well as to characterize the two SC states.

 \section{Acknowledgments}
  We thank K. Rana for assistance in the initial stage of the NMR measurements and A. Sapkota for fruitful discussions.  The research was supported by the U.S. Department of Energy (DOE), Office of Basic Energy Sciences, Division of Materials Sciences and Engineering. Ames National Laboratory is operated for the U.S. DOE by Iowa State University under Contract No.~DE-AC02-07CH11358. 

 \section{Data Availability}
 The data supporting the findings presented in this work are openly available \cite{Data}.

\clearpage
\begin{center}
{\bf Supplemental Material} 
\end{center}

\subsection{Hyperfine coupling tensors and internal fields at the Te sites}

\begin{figure*}
\centering
\includegraphics[width=\textwidth]{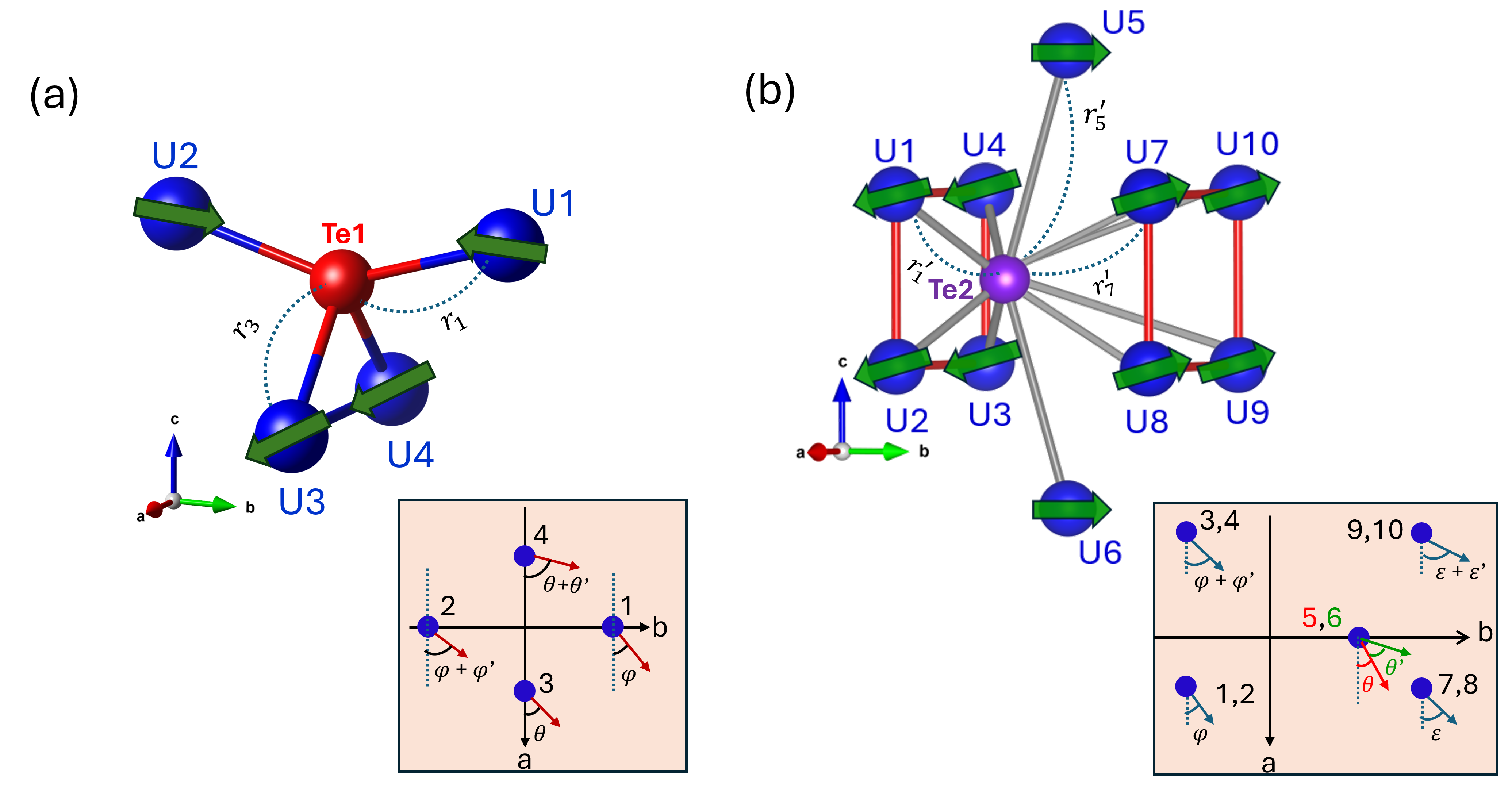}
\caption{Local configuration of the (a) Te1 and (b) Te2 sites. The blue circles represent the nearest-neighbor U atoms for each Te site. $r_i$'s, and $r'_i$'s denote the distance between U and Te site for the Te1 and Te2 sites, respectively.  Lower figures in (a) and (b) are the top view (from the  $c$ axis) of the magnetic-moment directions on each U ions specified by the angles $\phi,  \phi', \theta$, and $\theta' $ for the case of Te1 and $\phi$,  $\phi'$ , $\theta$, $\theta'$, $\epsilon$, $\epsilon'$ for the case of Te2.
 }
\label{fig:Fig-S1}
\end{figure*}

    One of the most striking features in the $1/T_1T$ data is that the Te(I) site detects the stronger antiferromagnetic (AFM) spin fluctuations along the $c$ axis compared to the Te(II) site. 
      These magnetic fluctuations at the Te sites arise from the hyperfine interactions between the $^{125}$Te nucleus and the U spins.  
     Therefore, it is crucial to understand the relationship between the directions of the hyperfine field at the Te sites and the U moments.
   The hyperfine field at the Te sites is due to transferred hyperfine interactions and can be described as the sum of contributions from the nearest-neighbor U moments:
\begin{equation}
H_{\rm hf} = \sum_{i=1}^{n} \mathbf {A_i \cdot m_i},
\label{Eq6}
\end{equation}
where \textit{n} is the number of nearest neighbor U atoms, \textit{\textbf{m$_i$}} is the magnetic moment at the \textit{i}-th U site and \textit{\textbf{A$_i$}} is the hyperfine coupling tensor between Te nucleus and \textit{i}-th U site. 
   
In the case of Te1 having a local symmetry of $mm2$, as shown in  Fig. S\ref{fig:Fig-S1} (a), there are four neighboring U atoms belonging to three different ladders, and Te1 is located at the center of the distorted tetrahedron formed by the U atoms. 
It is noted that the Te1 lies on two mirror planes: one perpendicular to the $a$ axis (which includes the U1 and U2 atoms) and another perpendicular to the $b$ axis (which includes the U3 and U4 atoms).

Given this symmetry, the components of the hyperfine coupling tensor  \textit{\textbf{A$_i$}} can be expressed  as
\begin{equation}
\mathbf {A_1} = \begin{pmatrix}
A_{aa} &  0   & 0\\
0 & A_{bb} &  A_{bc}\\
0 & A_{cb} &  A_{cc}
\end{pmatrix},
\end{equation}

\begin{equation}
\mathbf {A_2} = \begin{pmatrix}
A_{aa} &  0   & 0\\
0 & A_{bb} &  -A_{bc}\\
0 & -A_{cb} &  A_{cc}
\end{pmatrix},
\end{equation}

\begin{equation}
\mathbf {A_3} = \begin{pmatrix}
A_{aa} & 0 & -A_{ac}\\
0 & A_{bb} & 0\\
-A_{ca} & 0 & A_{cc}
\end{pmatrix},
\end{equation}

and
\begin{equation}
\mathbf {A_4} = \begin{pmatrix}
A_{aa} & 0 & A_{ac}\\
0 & A_{bb} & 0\\
A_{ca} & 0 & A_{cc}
\end{pmatrix}.
\end{equation}
where we assumed  the magnitude of each component of $A_1$ and $A_2$ is equal to that of $A_3$ and $A_4$ for simplicity as the distance $r_1$ (= 3.1817 \AA)  between U1 (or U2) and Te1 is close to  $r_3$ (= 3.0553 \AA), the distance  between U3 (or U4) and Te1 \cite{Hutanu2020_SM}. 
A similar hyperfine-tensor analysis for the same local environment has been performed on other compounds such as  EuGa$_4$ \cite{Yogi2013_SM}, EuCo$_2$As$_2$ \cite{Ding2017_SM}, and EuCo$_2$P$_2$ \cite{Higa2017_SM}.  

In the case of Te2 (local symmetry $m2m$), as shown in  Fig. S\ref{fig:Fig-S1} (b), it was reported that the 10 neighboring U spins should be considered, including the second and third nearest neighbor U atoms \cite{Fujibayashi2023_SM}. 
However, we found that including those more distant U sites does not alter the quantitative aspects of the following discussion.
Therefore, for simplicity, we consider only the four nearest-neighbor U atoms.
By considering the symmetry again,  the components of the hyperfine coupling tensor for the Te2 site \textit{\textbf{A'$_i$}} are described as  

\begin{equation}
\mathbf {A'_1} = \begin{pmatrix}
A'_{aa} &  - A'_{ab}  &  A'_{ac}\\
-A'_{ba} & A'_{bb} & -A'_{bc}\\
A'_{ca} & -A'_{cb} &  A'_{cc}
\end{pmatrix},
\end{equation}

\begin{equation}
\mathbf {A'_2} = \begin{pmatrix}
A'_{aa} &  - A'_{ab}  &  -A'_{ac}\\
-A'_{ba} &A'_{bb} & A'_{bc}\\
-A'_{ca} & A'_{cb} &  A'_{cc}
\end{pmatrix},
\end{equation}

\begin{equation}
\mathbf {A'_3} = \begin{pmatrix}
A'_{aa} &  A'_{ab}  &  A'_{ac}\\
A'_{ba} & A'_{bb} & A'_{bc}\\
A'_{ca} & A'_{cb} &  A'_{cc}
\end{pmatrix},
\end{equation}

and
\begin{equation}
\mathbf {A'_4} = \begin{pmatrix}
A'_{aa} &  A'_{ab}  & - A'_{ac}\\
A'_{ba} & A'_{bb} & -A'_{bc}\\
-A'_{ca} & -A'_{cb} &  A'_{cc}
\end{pmatrix}.
\end{equation}

Utilizing those hyperfine coupling tensors, we present the hyperfine field $H_{\rm hf}$ for each Te site. 
For simplicity, we assume that the U moments lie within the $ab$ plane, given that the $c$ axis is the magnetic hard axis under pressures \cite{Li2021}.
Here the direction of the magnetic moments $m$  for each U atom around the Te1 site are defined by four angles:  $\phi,  \phi', \theta$, and $\theta' $  as shown in Fig. S\ref{fig:Fig-S1} (a): 
$$
\begin{array}{cc}
m_1 =(m \cos{\phi}, m \sin{\phi}, 0), \\
\\
m_2 =(m \cos{(\phi + \phi')}, m \sin{(\phi + \phi')}, 0), \\
\\
m_3 =(m \cos{\theta}, m \sin{\theta}, 0), \\
\\
m_4 =(m \cos{(\theta + \theta')}, m \sin{(\theta + \theta')}, 0).\\
\\
\end{array}
$$
Under these assumptions, ${\bf H}_{\rm int}$(Te1) at Te1 can be expressed as: 
$\mathbf {H_{i\rm nt}}$(Te1)    =  
\begin{equation}
 \begin{pmatrix}
A_{aa} m [\cos{\phi}+\cos({\phi+\phi'}) + \cos{\theta}+\cos({\theta+\theta'})]\\
A_{bb} m [\sin{\phi}+\sin({\phi+\phi'}) +  \sin{\theta}+\sin({\theta+\theta'})]\\
A_{cb} m[\sin{\phi} - \sin({\phi+\phi'})] - A_{ca} m[\cos{\theta}-\cos({\theta+\theta'})]
\end{pmatrix}.
\end{equation}

We also derive the ${\bf H}_{\rm int}$(Te2) at the Te2 site. 
Again, assuming the U magnetic moments are in the $ab$ plane as shown in  Fig. S\ref{fig:Fig-S1} (b) with 
$ m_1 =m_2 =(m \cos{\phi}, m \sin{\phi}, 0)$ and $m_3 =m_4 =(m \cos{(\phi + \phi')}, m \sin{(\phi + \phi')}, 0), $
${\bf H}_{\rm int}$(Te2) is calculated to be \\
$\mathbf {H_{i\rm nt}}$(Te2)    =  
\begin{equation}
\begin{pmatrix}
2A'_{aa} m [\cos{\phi}+\cos({\phi+\phi'})] - 2A'_{ab} m [\sin{\phi}-\sin({\phi+\phi'})]\\
2A'_{bb} m [\sin{\phi}+\sin({\phi+\phi'})] - 2A'_{ba} m [\cos{\phi}-\cos({\phi+\phi'})]  \\
0
\end{pmatrix}.
\end{equation}

Interestingly, the $\mathbf {H_{i\rm nt}}$ for Te1 has the finite $c$ axis component while no $c$ axis component in $\mathbf {H_{i\rm nt}}$ for Te2 when the U spins are in the $ab$ plane.
This difference may lead to the different hyperfine field fluctuations along the $c$ axis at the two Te sites, potentially explaining the difference in $R_{i,{\rm c}}$ for Te(I) and Te(II).  
As $R_{i,{\rm c}}$ for Te(I) enhances more than that of Te(II) at low temperatures shown in the main text, it is most likely that Te(I) and Te(II) are assigned to the Te1 and Te2, respectively, in the crystal structure.
This is the same with the initial assignment of Te NMR lines in ref. \cite{Tokunaga2019_SM},  and Fujibayashi \textit{et al}. also pointed out such a possibility in ref. \cite{Fujibayashi2023_SM}.
 
It is also interesting to mention that we find a small contribution from U5 and U6 to the $c$ components in the  ${\bf H}_{\rm int}$(Te2)  if we include the second and third nearest neighbor. 
This could explain the small enhancement in $R_{i,{\rm c}}$ for Te(II) at low temperatures under a pressure of 1.05 GPa. 
However, it should be noted that, as pointed out in the main text, Te(I) also detects the AFM spin fluctuations along the $a$ axis even though ferromagnetic fluctuations are dominated along this axis. This cannot be simply explained by the model, requiring more detail analysis for hyperfine coupling tensors. 
For example, for Te(I), we assume a perfect tetrahedral configuration for simplicity, even though it is slightly distorted. The small distortion may produce an additional component in the hyperfine tensor along the $a$ axis. Similarly, for Te(II), we also simplified the calculation by considering only four nearest-neighbor U ions. Those simplicities may produce small deviations from the actual hyperfine coupling tensors for both Te sites, which may be responsible for the reason why AFM spin fluctuations are picked up along the $a$ axis.

\begin{figure*}
\centering
\includegraphics[width=0.9\textwidth]{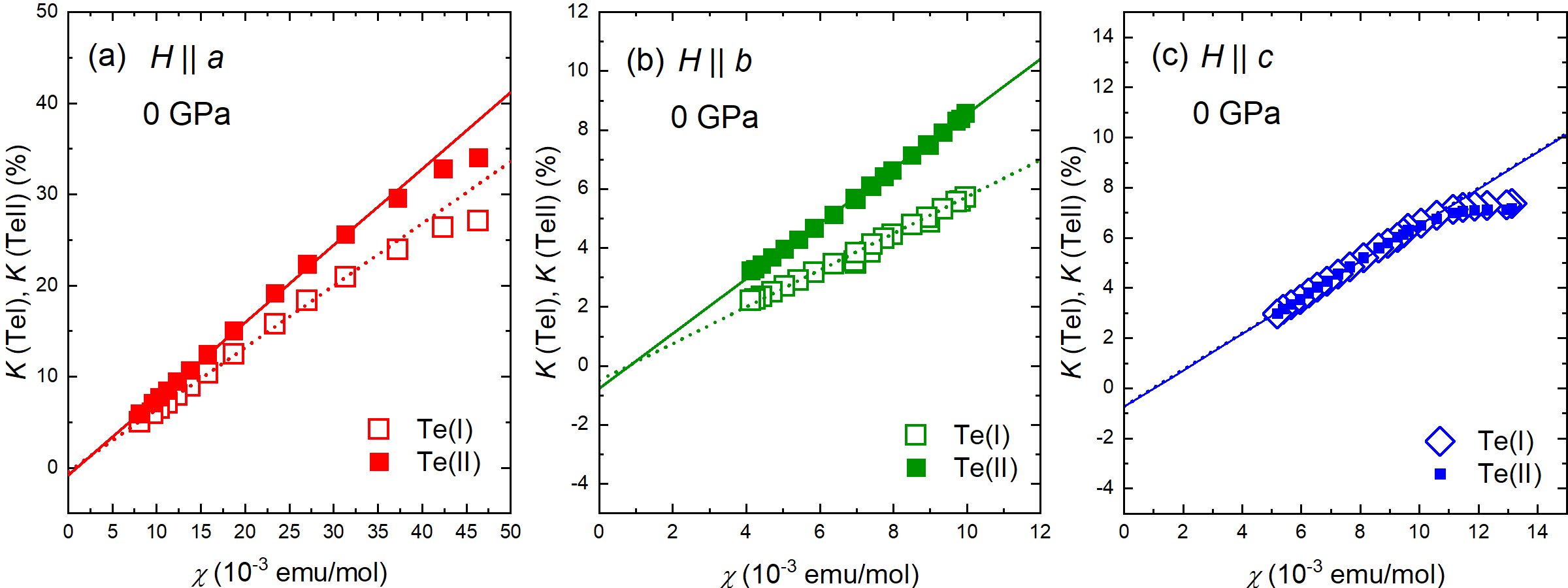}
\caption{$K$ versus $\chi$ plot at ambient pressure with temperature as implicit parameter for magnetic field along (a) $a$ axis, (b) $b$ axis, and (c) $c$ axis. Here data for Te(I) is shown in open symbols and Te(II) in solid symbols. The linear fits of each site are given by dotted and solid lines for Te(I) and Te(II) sites, respectively.}
\label{fig:Fig-S2}
\end{figure*}

 Finally, it is worth considering the internal fields at the Te sites above $p_{\rm c}$.
 As described in the main text, the signal intensity of Te(II) relative to Te(I) decreases below $\sim$ 8-15 K (depending on $p$), likely due to the short-range magnetic ordering.
It would be reasonable to assume that the ferromagnetic correlation develops first within the U ladders during the short-range magnetic ordering, followed by the antiferromagnetic correlation between the U ladders.   
   Since the Te2 site has four nearest neighbor U moments from a single U ladder,  while the internal field at the Te1 site is influenced by the four U moments from three U ladders, one might anticipate a greater hyperfine field at Te2 in the short-range ordered state. 
   Based on the temperature dependence of the signal intensity ratio of Te(II) to Te(I), Te(I)and Te(II) could be assigned to Te1, and Te2, respectively, consistent with the site assignment described above.

\subsection{Knight Shift $K$ versus magnetic susceptibility $\chi$ }

To obtain the hyperfine coupling constants which we reported in Sec. III. A of the main text, we used the relationship between the Knight shift $K$ and the magnetic susceptibility $\chi$ given as:
\begin{equation}
K(T) = K_0 + \frac{A_{\mathrm{hf}}}{N_{\mathrm{A}}\mu_{\mathrm{B}}}\chi(T),
\end{equation} 
where $K_0$ is the temperature-independent part of the Knight shift, $N_{\mathrm{A}}$ is the Avogadro number, and $\mu_{\mathrm{B}}$ is Bohr magneton. Here, $A_{\mathrm{hf}}$ is the hyperfine coupling constant between the $^{125}$Te nucleus and the U spins, and from the slope of the linear fit of the $K$-$\chi$ plot (see Fig. S\ref{fig:Fig-S2}), we can estimate the value of $A_{\mathrm{hf}}$.  The hyperfine coupling constants are estimated to be $A_{aa}$(Te(I)) = 3.8 T/$\mu_{\rm B}$ and  $A_{aa}$(Te(II)) = 4.7 T/$\mu_{\rm B}$ for the $a$ axis, $A_{bb}$(Te(I)) = 3.5 T/$\mu_{\rm B}$ and  $A_{bb}$(Te(II)) = 5.2 T/$\mu_{\rm B}$ for the $b$ axis, and $A_{cc}$ = 4.1 T/$\mu_{\rm B}$ for both Te sites under magnetic field parallel to the $c$ axis. The small deviation from the linear behavior observed in the $K$-$\chi$ plot for field along the $a$ and $c$ axes below $\sim 10$ K  could be attributed to the U atom defects in the low-$T_\mathrm{c}$ crystals as discussed in the main text.

\subsection{$R_{i,\alpha} / K$ versus $T$ }

\begin{figure*}
\centering
\includegraphics[width=0.9\textwidth]{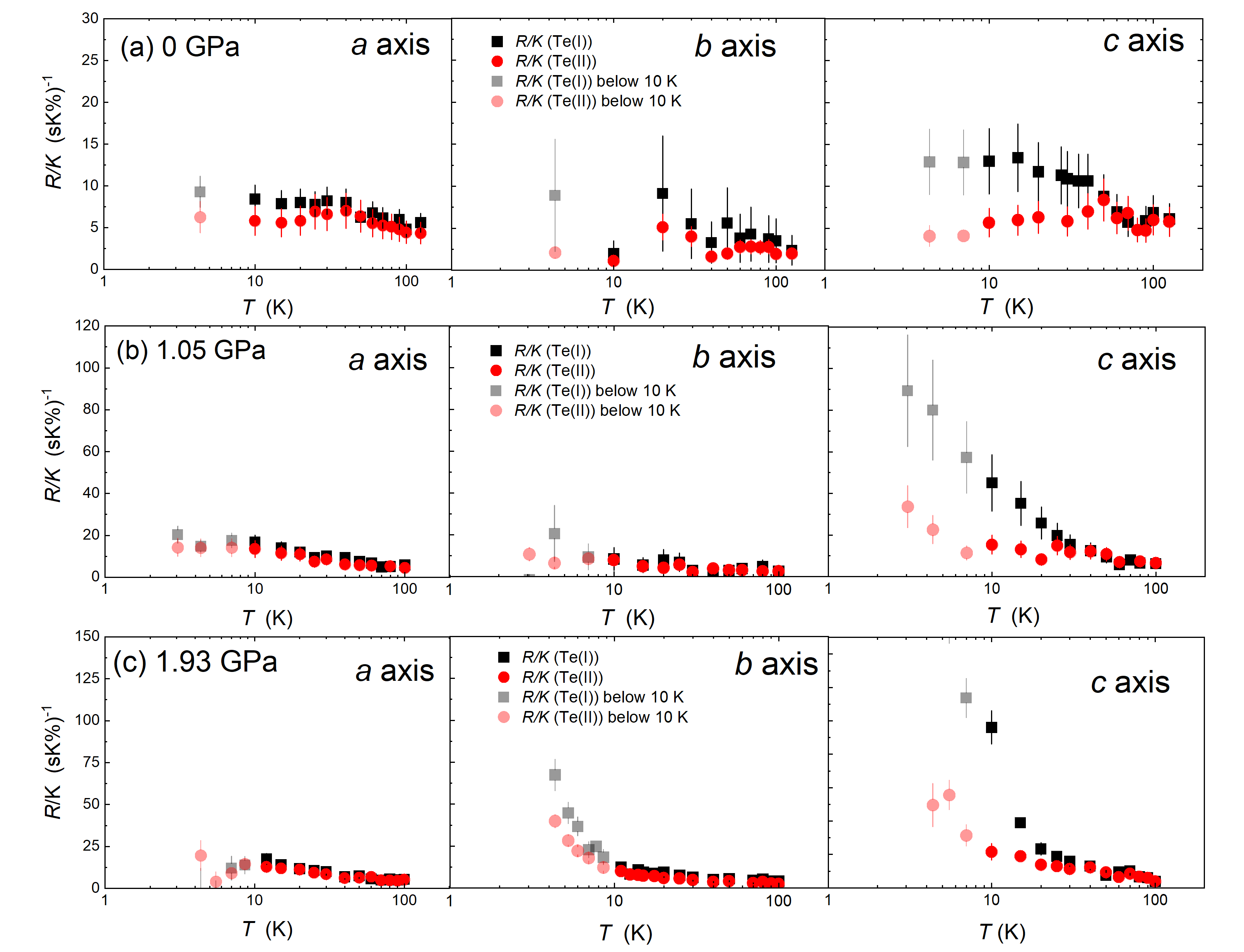}
\caption{Temperature dependence of $R_{a} / K_{a}$ (left), $R_{b} / K_{b}$ (middle), and $R_{c} / K_{c}$ (right) for the $p$ = 0 GPa (a), 1.05 GPa (b), and 1.93 GPa (c) for Te(I)  (black squares) and Te(II) (red circles). The $R_{i,\alpha} / K$ data below 10 K were also shown with translucent symbols for reference.}
\label{fig:Fig-S3}
\end{figure*}

In the main text, we derived the temperature dependence of the directional magnetic fluctuation components, $R_{i,\alpha}$, where $i =$ I, II and $\alpha, \beta, \gamma $ = $\{a, b, c\}$ for each Te($i$) site. 
By comparing the temperature dependence of $R_{i,\alpha}$ with that of the corresponding $K$, we discussed the nature of these fluctuations in the main text.  
Here we show the temperature dependence of  $R_{i,\alpha}$ divided by the corresponding $K$ values for the two Te sites under three different pressures in Fig. S\ref{fig:Fig-S3}, which is another way to determine whether $R_{i,\alpha}$ scales with $K$ or not.

 At ambient pressure,  $R_{i,c} / K_c$ for Te(I) site becomes greater than that of $R_{i,c} / K_c$ for Te(II) at low temperatures. This clearly indicates the Te(I) site detects additional AFM magnetic fluctuations along the $c$ axis, as we discussed in the main text.  
 As for the $a$ axis, although $R_{i,a} / K_a$ exhibits a similar behavior for both the Te sites, it is clear that $R_{i,a} / K_a$ for Te(I) is slightly greater than that of Te(II) at low temperatures (see Fig. S\ref{fig:Fig-S3} (a)). 
 This suggests that Te(I) picks up additional $q \neq 0$ fluctuations, evidencing the finite AFM fluctuations along the $a$ axis even at ambient pressure, as pointed out in the main text, although ferromagnetic $q = 0$ fluctuations are dominated along the $a$ axis.  
  Along the $b$ axis, the temperature dependence of $R_{i,b} / K_b$ is nearly site-independent within the measurement uncertainty.

As pressure is increased, the enhancement of $R_{i,c} / K_c$ for the Te(I) site compared to the Te(II) site becomes more evident, proving again that Te(I) is more sensitive to AFM fluctuations than Te(II). However, the site difference in the magnetic fluctuations observed at ambient pressure along the $a$ axis becomes less clear with an increase in pressure (see Fig. S\ref{fig:Fig-S3} (b) and (c)). Above $p_{\rm{c}}$, we observed a significant enhancement of AFM spin fluctuations along the $b$ axis; however, no profound difference is seen between the two Te sites.

\subsection{$^{125}$Te spin-lattice relaxation rate 1/$T_1$ under $H$ $\parallel$ $b$ at $p$ = 2.05 GPa }

\begin{figure*}
\centering
\includegraphics[width=0.5\textwidth]{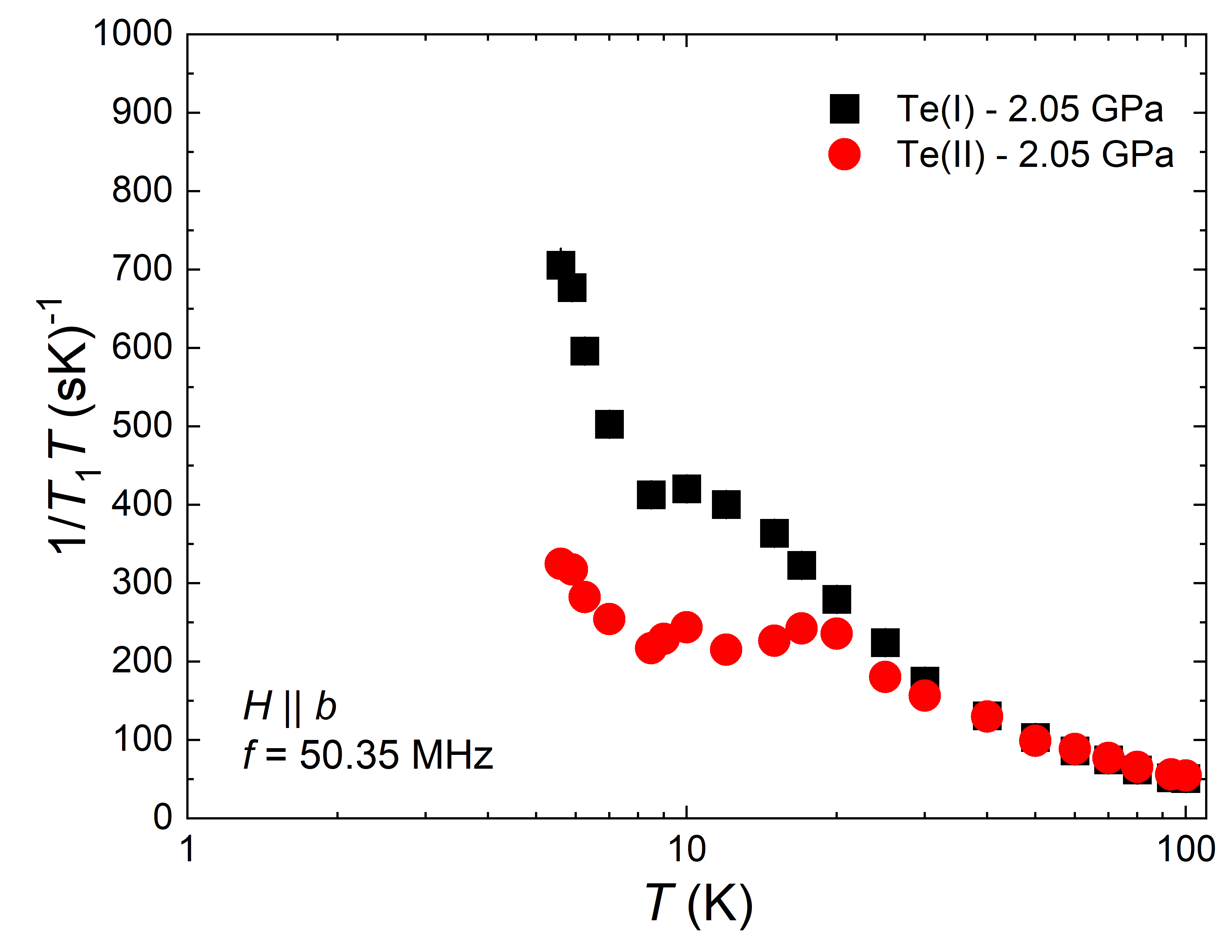}
\caption{1/$T_1T$ data for both Te(I) site (black squares) and  Te(II) site (red circles) under $H$ $\parallel$ $b$ at $p$ = 2.05 GPa measured at the frequency $f = 50.35$ MHz.}
\label{fig:Fig-S4}
\end{figure*}

Fig. S\ref{fig:Fig-S4} shows the temperature dependence of  1/$T_1T$ data for both Te sites under $H$ $\parallel$ $b$ at $p$ = 2.05 GPa measured at the frequency $f = 50.35$ MHz.  The figure clearly features the additional enhancement of 1/$T_1T$ for Te(I) compared to Te(II) under pressure. This difference is used to estimate (1/$T_1T$)$_{\rm{AFM}}$, a tentative quantity to represent AFM fluctuation at 2.05 GPa in the main text.


\end{document}